\documentclass[aps,prd,preprint,superscriptaddress]{revtex4}
\usepackage{epsfig}
\usepackage{amssymb}
\usepackage{amsfonts}
\usepackage{subfigure}
\usepackage{float}
\begin{document}

\preprint{HU-EP-03/43}
\preprint{IFUP-TH-2003/26} 
\preprint{TRINLAT-03-02} 
\preprint{$--$ Revised Version $--$}

\title{Universality, vortices and confinement: 
\\ modified SO(3) lattice gauge theory at non-zero temperature}
\author{A.~Barresi} 
\affiliation{Dipartimento di Fisica e INFN, Pisa, Italy}
\author{G.~Burgio}
\affiliation{School of Mathematics, Trinity College, Dublin 2, Eire}
\author{M.~M\"uller-Preussker}
\affiliation{Humboldt-Universit\"at zu Berlin, Institut f\"ur Physik, 
Berlin, Germany}
\date{\today}
\begin{abstract}
We investigate the adjoint $SU(2)$ lattice gauge theory in 3+1 
dimensions with the Wilson plaquette action modified by a 
$\mathbb{Z}_2$ monopole suppression term. 
For the zero-twist sector we report on indications for the existence of 
a finite temperature transition decoupled from the unphysical 
bulk transitions. 
\end{abstract}
\maketitle


\section{Introduction}
Pure 3+1 dimensional $SU(N)$ lattice gauge theories in the fundamental 
representation show a finite temperature deconfinement phase transition
\cite{McLerran:1981pk,Kuti:1981gh} 
together with the spontaneous breaking of a global $\mathbb{Z}_N$ center
symmetry \cite{Polyakov:1978vu,Susskind:1979up} 
governing the critical indices, which e.g. for $SU(2)$ correspond to the 
universality class of the 3D Ising model \cite{Svetitsky:1982gs}. 
Lattice universality arguments are commonly 
used to claim that the same should hold for 
any possible lattice action discretization, in particular with different
gauge group representations. On the other hand, if confinement of quarks 
can be ascribed to peculiar (topological) excitations of the continuum 
Yang-Mills fields, it is not clear how center symmetry breaking can 
lead to an effective theory of QCD \cite{Fortunato:2000vf}.
At finite temperature vortices classified 
through $\pi_1(SU(N)/\mathbb{Z}_N)\sim\mathbb{Z}_N$ 
along a compactified dimension can of course provide a 
suitable degree of freedom for the pure Yang-Mills case \cite{'tHooft:1979uj}, 
but their fate in the presence of fermions remains challenging.
Moreover, in SUSY Yang-Mills theory \cite{Seiberg:1994rs,Seiberg:1994aj} 
and  in the Georgi-Glashow model \cite{Polyakov:1977fu}
confinement is driven by magnetic monopoles.
Recent work regarding Yang-Mills theories based on exceptional
groups \cite{Holland:2003jy} conjectures that vortices might not be at all 
necessary to have a confining theory. 
What r\^{o}le do vortices play then in $SU(N)$?
An investigation of the $SU(2)$ lattice gauge theory in the
{\it center blind} adjoint representation, i.e. 
$SO(3)\sim SU(2)/\mathbb{Z}_2$, might offer some interesting insight. 
The presence of  finite temperature effects
for such a model has been debated for a long time. 
Two main problems have been faced.
First of all Polyakov's center symmetry breaking mechanism is available 
only for the half-integer representations. 
For integer ones the $\mathbb{Z}_2$ local invariance makes
the fundamental Polyakov loop of no use.
The adjoint Polyakov loop, on the other hand, \emph{cannot} be strictly
speaking an order parameter for a transition, assuming it existed, 
since gluons will screen adjoint quarks at some distance
($\sim 1.25~\mathrm{fm}$ for $SU(2)$)
\cite{Michael:1985ne,deForcrand:1999kr}. 
However, it can still be taken as a signature 
to distinguish, at finite temperature, a confined although
screened phase from a deconfined one. Anyhow, the problem of establishing a
well defined criterion for confinement in this case remains an open and 
interesting one. 

Second, and most importantly,
lattice artifacts lead to first order bulk phase
transitions at strong coupling, preventing the continuum limit to be reached 
within the ordinary confined phase \cite{Bhanot:1981eb,Greensite:1981hw}. 
Similar phase diagrams are shared by $SU(N)$ theories with 
$N\ge 3$ \cite{Creutz:1982ga}. For $SU(2)$ $\mathbb{Z}_2$ monopoles were 
observed to drive the bulk transition \cite{Halliday:1981te,Caneschi:1982ik}. 
Another interesting development came as the vortex free energy, 
measured on the lattice already in the fundamental theory 
\cite{deForcrand:2001nd}, was recently 
computed also in the adjoint theory \cite{deForcrand:2002vs},
thus suggesting how center vortices
and twist sectors are entering a center-blind theory.
The latter work, however, falls short of any attempt to investigate the 
theory at finite temperature, stumbling on ergodicity problems 
of the algorithm at the bulk phase transition, where tunnelling among 
different twist sectors becomes strongly suppressed. 
Thus, a thorough finite temperature investigation
of the $SO(3)$ theory taking into account the topological excitations
is still missing. Several attempts searching for a
decoupling of the finite temperature transition from the bulk transitions
were originally undertaken by S. Datta and R. Gavai 
(see \cite{Datta:1998nv,Datta:1999np,Datta:1999ep} and further citations 
therein). These authors used the fundamental-adjoint mixed action modified
by $\mathbb{Z}_2$ monopole and vortex suppression terms within the
Villain-type formulation. 
By studying the specific heat 
a finite temperature transition consistent with the Ising universality class
was found. In the pure center-blind adjoint case including only 
monopole suppression it could however only be established for the smallest 
time-extension $N_t=2$. Moreover, the relevance of different twist sectors 
had not yet appeared in the literature.   

Here we want to go a step further employing Wilson's adjoint action 
formulation modified by a similar $\mathbb{Z}_2$ monopole suppression. 
In this case the action itself is manifestly center-blind.
The corresponding phase structure looks in many respects similar 
to the Villain case,  but it differs - as we shall show - substantially 
in the fundamental-adjoint coupling plane. The areas along the two axes are 
completely separated by a bulk phase transition and thus the proof of 
universality - if the latter is really fulfilled - becomes more complicated. 
In our investigation the twist variables will prove an important 
ingredient to understand the phase structure of the model.  

The outline of the paper is as follows. In Section \ref{sec2} we introduce
the model and discuss its phase structure at vanishing temperature with 
standard bulk observables. In Section \ref{sec3} twist variables will be
used in order to show that the first oder bulk transition seems to weaken to
second order in a certain $\beta_A - \lambda$ range. 
In Section \ref{sec4} we report on the existence of the 
finite temperature transition and check scaling for the critical temperature.
Section \ref{sec5} contains our conclusions.
Reports of this work at early stages have been published in 
\cite{Barresi:2001dt,Barresi:2002un}.

\section{Adjoint action model with $\mathbb{Z}_2$ monopole suppression} 
\label{sec2}
We study the $SU(2)$ lattice gauge theory with a mixed 
fundamental-adjoint representation Wilson action
modified by a chemical potential term suppressing $\mathbb{Z}_2$ monopoles
\begin{eqnarray} 
S=\sum_{P} \Bigg[
  \frac{4}{3}\beta_{A} \left(1-\frac{\mathrm{Tr}_{F}^{2}U_{P}}{4}\right)
  + \beta_{F} \left(1-\frac{\mathrm{Tr}_{F}U_{P}}{2}\right) \! \Bigg]
  +\lambda \sum_{c}(1-\sigma_{c})\,, \\
  \frac{1}{g^2} = \frac{\beta_F}{4}+2\frac{\beta_A}{3} \nonumber
\label{ouraction}
\end{eqnarray}
where $\sigma_{c}=\prod_{P\in\partial c}\mathrm{sign}(\mathrm{Tr}_{F}U_{P})$
taken as a product around elementary 3-cubes $~c~$ defines the 
$\mathbb{Z}_2$-charge. 
For these monopoles a density can be introduced as 
\begin{eqnarray} \label{monopdens}
M = 1-\langle\frac{1}{N_c}\sum_{c}\sigma_{c}\rangle 
\end{eqnarray}
\noindent
normalized such that it tends to unity in the strong coupling region 
($N_c$ denotes the number of 3-cubes on the 4D lattice). 
We will be particularly interested in the pure adjoint theory, i.e. 
$\beta_{F}=0$.
In the latter case one can analyze the model with the link variables
represented both by $SO(3)$ or $SU(2)$ matrices, exploiting the property
$\mathrm{Tr}_{A}= \mathrm{Tr}_{F}^2-1$ for the Wilson term or picking
a random $SU(2)$ representative of the $SO(3)$ link to construct 
the $\mathbb{Z}_2$ monopole contribution. 
As expected nothing changes in the phase diagram, the integration over the
fundamental links simply doubling the integration domain in the partition 
function.
A standard Metropolis algorithm has been used to update the links in both 
cases and we use either one according to the best performance in the case
at hand. Although $~\sigma_{c}~$  is constructed in terms of fundamental 
representation plaquettes, it is a {\it natural}
$SO(3)$ quantity. In fact, for every given $SO(3)$ link variable,
the corresponding $SU(2)$ representative can always be determined
up to a sign. But the latter is cancelled in the product of plaquettes 
around any 3-cube, since each link occurs twice for each edge. 
In other words
\begin{equation}
U_{\mu}(x) \rightarrow -U_{\mu}(x) \quad \Rightarrow \quad
\sigma_{c}\rightarrow\sigma_{c}\,, \qquad \forall \mu,x,c \,.
\label{so3_inv}
\end{equation}
This ensures that for $\beta_F=0$ the action we study is center-blind
in the entire $\beta_A-\lambda$ plane.

\begin{figure}[htb]
\begin{center}
\includegraphics[angle=0,width=0.45\textwidth]{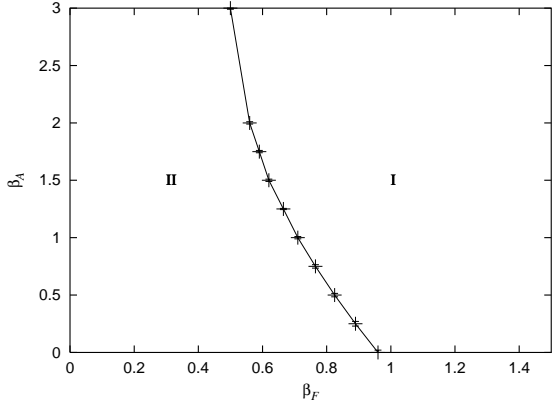}
\end{center}
\caption{The bulk phase transition of the mixed fundamental-adjoint action 
model (\ref{ouraction}) with monopole suppression for $\lambda=1.0.$}
\label{bfbalm}
\end{figure}
Previous finite temperature investigations were mainly done in the
Villain discretization for the $SO(3)$ term in the action,
introducing an independent $\mathbb{Z}_2$-valued
pla\-quette variable  $~\sigma_{P}~$ 
\cite{Halliday:1981tm,Datta:1999np,Datta:1999ep}. 
Two terms with chemical potentials were added in order
to suppress completely the effect of the lattice artifacts,
$\mathbb{Z}_2$  monopoles and vortices, whose densities in this case
are given by
\begin{eqnarray} \label{gavai}
\widetilde{M}&=&1-\langle\frac{1}{N_{c}}\sum_{c}\widetilde{\sigma}_{c}\rangle\,, \qquad
\widetilde{\sigma}_{c}=\prod_{P\in\partial c}\sigma_{P}\,,  \\
\widetilde{E}&=&1-\langle\frac{1}{N_{l}}\sum_{l}\widetilde{\sigma}_{l}\rangle\,, \qquad
\widetilde{\sigma}_{l}=\prod_{P\in\hat{\partial} l}\sigma_{P}\,, 
\end{eqnarray}
where $c$ and $l$ label the $N_c$ 3-cubes and the $N_l$ pro\-ducts of 
plaquettes having a link in common, respectively.
We can see that the monopole suppression term 
in Eq. (\ref{ouraction}) looks formally
identical with the one used in connection with the Villain-type action, but
its realization is different and leads to a different phase structure. 
The Villain discretization (on $\mathbb{R}^4$) can be proven to be equivalent
to the standard {\it fundamental} Wilson action for $\lambda_V \rightarrow \infty$ 
\cite{Halliday:1981tm,Datta:1999np,Datta:1999ep} (see Eq. (\ref{twadj})).
On the other hand one can show that the limit $\lambda\rightarrow\infty$ 
of the Wilson adjoint formulation can be mapped,
in the trivial twist sector, to some positive plaquette model 
\footnote{This equivalence has been first brought to our attention by O.~Jahn.}.
Such a mapping however, as we shall see below, is not equivalent to a 
continuous connection with the standard fundamental action theory. A simple 
inspection of the $\beta_{A(V)} -\beta_F$ phase diagram in these two 
limits shows that the differences are conspicous. Therefore,
$\lambda_V$ and $\lambda$ must not simply be identified
\cite{Barresi:2001dt,Barresi:2002un,deForcrand:2002vs}. One
should always bear in mind that in the Villain case the $SO(3)$ 
invariance under $U_{\mu}(x) \rightarrow -U_{\mu}(x)$ is not realized
in the action itself but only once the auxiliary $\sigma_P$ variables 
are integrated out. 
In Fig. \ref{bfbalm} the $\beta_F-\beta_A$ phase diagram of our model 
(\ref{ouraction}) at $T=0$ is shown for rather strong monopole suppression 
$\lambda=1.0$. Phase I is connected with the ordinary confinement phase
of the standard $SU(2)$ model within the fundamental representation, whereas
phase II extending to $\beta_A \to \infty$ occurs completely decoupled from
phase I for $\beta_A \ge 0$. Indeed, simulating the model with the lattice size 
up to $12^4$ the average plaquette variable 
exhibits a strong discontinuity across and metastable states on top of the 
transition line. At $\beta_A=0.0$ the latter was located at 
$\beta_F \simeq 0.96$. 
By studying the fundamental representation Polyakov loop and its susceptibility 
for lattice size $4 \times 12^3$ the finite-temperature phase transition 
was seen at $\beta_F \simeq 1.35$, i.e. within phase I as one would expect.  
The phase structure clearly differs from that of the Villain case, where 
the transition line for increasing $\beta_F$ extends along the latter axis 
up to $\beta_F \to \infty$ (see \cite{Datta:1999ep}). Therefore,
universality of physics observed for $\beta_F=0$ in comparison with
$\beta_A=0$ remains an open question.

\begin{figure}[htb]
\begin{center}
\includegraphics[angle=0,width=8cm]{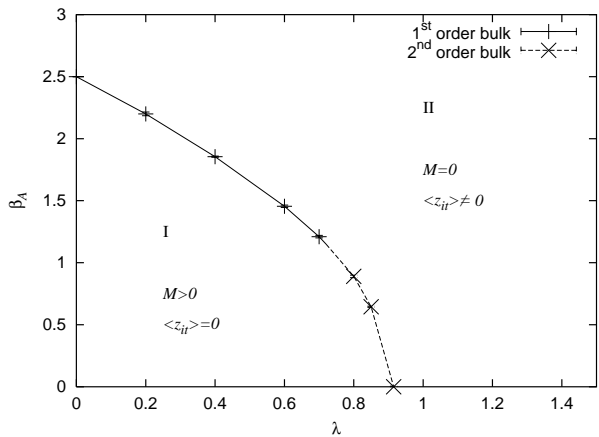}
\end{center}
\caption{The bulk phase transition in the $\beta_A-\lambda$ plane as seen
for lattice size $4 \times 12^3$.}
\label{bala}
\end{figure}
For $\beta_F=0$ -- as one turns on the monopole suppression --
the bulk transition moves down from $\beta_A \simeq 2.5$ to lower 
$\beta_A$-values intersecting the $\lambda$-axis at $\lambda \simeq 0.92$
as one can see from Fig. \ref{bala}. The phases I and II are denoted as
in Fig. \ref{bfbalm}.  The bulk transition is characterized by the 
condensation of $\mathbb{Z}_2$ monopoles within phase I $~(M > 0)~$ 
and by their suppression in phase II $~(M=0)$. 
It has been located by monitoring the monopole 
density $M$ (Fig. \ref{monopole}), the plaquette and the 
twist variables (for the definition of the latter see below) as a function of
$\beta_A$ for varying $\lambda$. Fig. \ref{monopole} for a lattice size of
$4 \times 12^3$ (i.e. for non-zero temperature) shows the transition to be 
discontinuous for small $~\lambda~$ turning into a continuous one 
(probably second order) at larger $~\lambda$ and lower $\beta_A$,
respectively (see the next paragraph). The end point of the first-order
branch (at $\beta_A \simeq 1.2$ for $N_t=4$) seems to be identical with 
the lower $\lambda$ endpoint of the finite temperature transition to be 
reported in Section \ref{sec4}.  
\begin{figure}[ht]
\begin{center}
\subfigure[$\lambda=0.0$]{
\includegraphics[angle=0,width=0.36\textwidth]{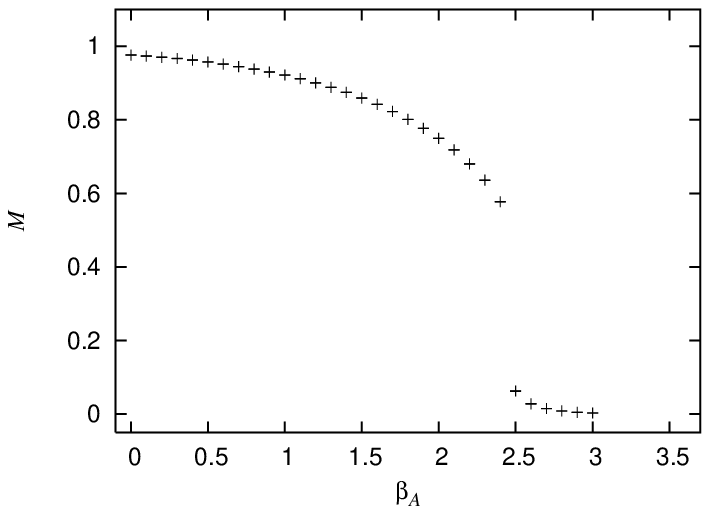}}
\subfigure[$\lambda=0.4$]{
\includegraphics[angle=0,width=0.36\textwidth]{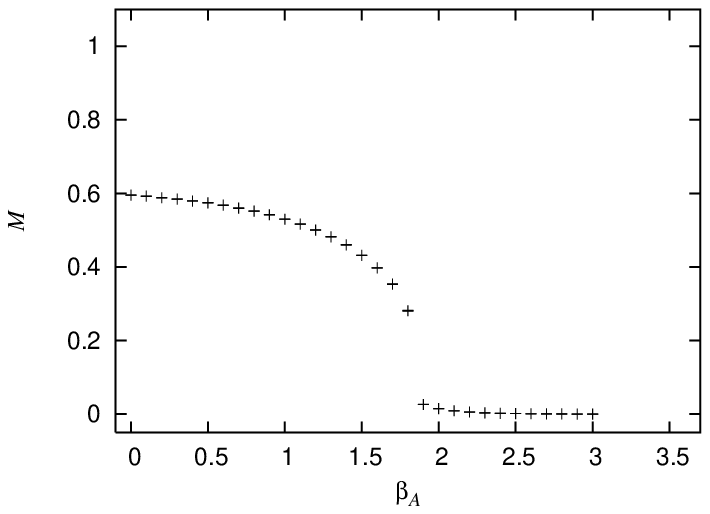}}
\subfigure[$\lambda=0.6$]{
\includegraphics[angle=0,width=0.36\textwidth]{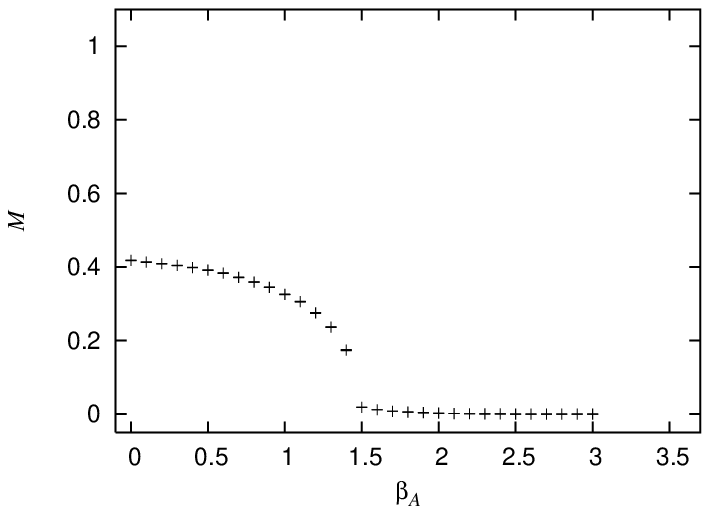}}
\subfigure[$\lambda=0.7$]{
\includegraphics[angle=0,width=0.36\textwidth]{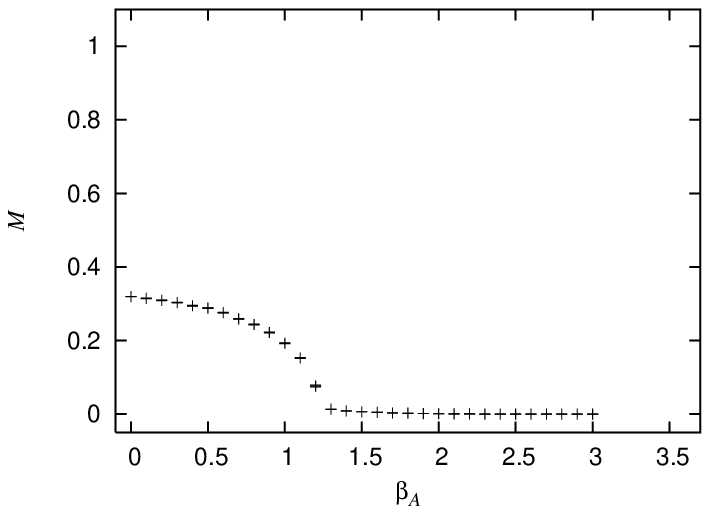}}
\subfigure[$\lambda=0.8$]{
\includegraphics[angle=0,width=0.36\textwidth]{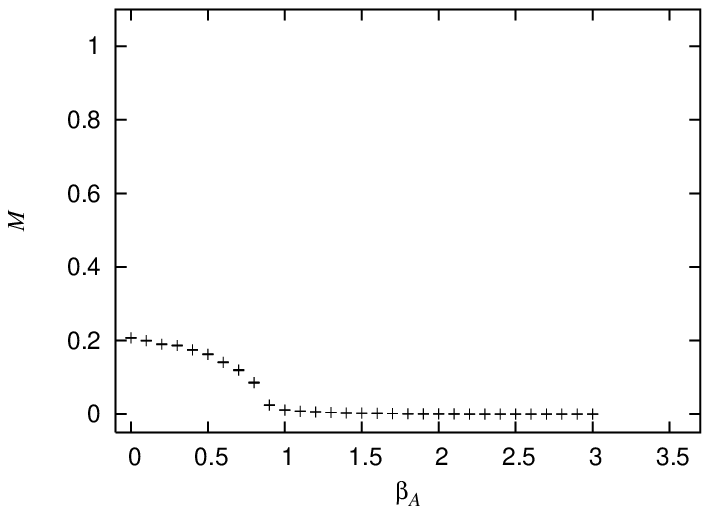}}
\subfigure[$\lambda=1.0$]{
\includegraphics[angle=0,width=0.36\textwidth]{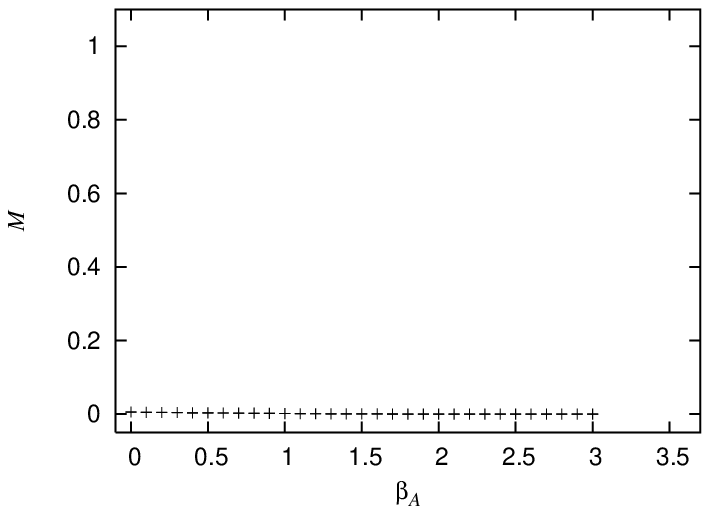}}
\end{center}
\caption{Monopole density as a function of $\beta_A$ for different
values of $\lambda$ ($V=4\times 12^3$) with a statistics
of $O(10^5)$ configurations.}
\label{monopole}
\end{figure}
For $~\beta_A >0,~\lambda \ge 1.0~$ no bulk transition is observed anymore.  
Our $\beta_A-\lambda$ phase diagram looks very similar to that 
of the Villain case in the $\beta_A-\lambda_V$ plane
\cite{Halliday:1981tm,Datta:1999np}. But note that in the Wilson case
phase II seems to be disconnected from phase I. 
 
In any case the properties of the different phases 
should be established by means of observables well defined in the
continuum theory, e.g. the glueball spectrum or the screening mass. 
Such a project is currently under consideration.

\section{Twist sectors and tunneling}
\label{sec3}
Vortex free energies related to 't Hooft spatial loops
were recently measured on the lattice within the
fundamental and the adjoint representation of $SU(2)$
\cite{deForcrand:2001nd,deForcrand:2002vs}. 
In the $SO(3)$ case the Villain discretization
was used and it was stressed that on the torus $T^4$
the following equivalence holds
\begin{eqnarray}
\sum_{\mathrm{t.s.}} Z_{SU(2)}
\equiv A \sum_{\sigma_P=\pm 1} &\int& (DU) 
e^{\beta_{V}\sum_{P}\sigma_{P}\mathrm{Tr}_{F}U_{P}}
\prod_c~\delta(\widetilde{\sigma}_c-1)\,,
\label{twadj}
\end{eqnarray}
where the left hand side is summed over all the twist sectors.
To obtain the equivalence of this modified $SO(3)$ Villain model
with the $SU(2)$ Wilson action case
with standard periodic boundary conditions the global constraint
\begin{eqnarray}
N_{\mu\nu}\equiv\prod_{P\;\in\; \mathrm{plane}\; \mu\nu}\sigma_P=+1\,.
\label{constr}
\end{eqnarray}
has to be satisfied. It is straightforward to see
that an observable able to distinguish between trivial
and non-trivial twist sectors is given by
\begin{eqnarray}
z_{\mu\nu}\equiv\frac{1}{L_\rho L_\sigma}\sum_{\rho\sigma}
                \prod_{P\;\in\; \mathrm{plane}\; \mu\nu}
\mathrm{sign Tr}_F U_P\,,
\hspace{.1cm}(\epsilon_{\rho\sigma\mu\nu}=1) .
\label{deftwi}
\end{eqnarray}
These {\it twist variables} are
again truly $SO(3)$ observables since due to the boundary conditions 
the signs of the 
links in the fundamental representation drop out in the product,
the plane extending over the whole length of the space and time 
directions. 

A connection between the existence of non-trivial twist sectors, 
the presence of center vortices and, in the deconfined phase, the 
occurence of a state characterized by a value
$-1/3$ for the adjoint Polyakov loop $~L_A~$ 
(first observed in \cite{Cheluvaraja:1996zn,Datta:1998nv})
was proposed by de~Forcrand and Jahn \cite{deForcrand:2002vs}.
The latter authors found the twist sectors to become
metastable under local updates close to the bulk transition, 
i.e. the suppression of $\mathbb{Z}_2$ monopoles causes the tunnelling
among different vortex vacua to become exponentially suppressed 
with increasing volume. The authors were thus prevented to go further
and to investigate the possible extension of such connection into the
finite temperature case.
\begin{figure}[htb]
\begin{center}
\includegraphics[angle=0,width=0.48\textwidth]{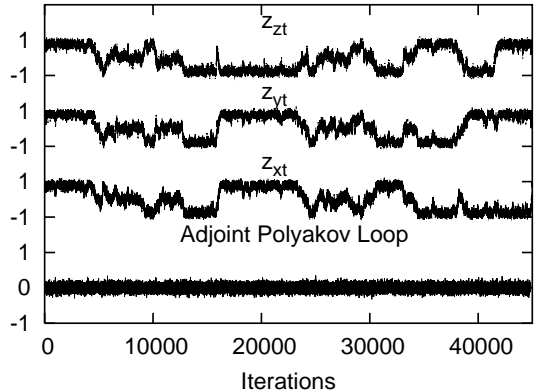}
\end{center}
\caption{Monte Carlo history of the adjoint Polyakov loop
and of the three electric twist variables at
$\lambda=0.858,\beta_A=0.65$ for $V=12^4$.}
\label{newtwist}
\end{figure}
In the Wilson case, turning on the $SO(3)$ invariant 
$\mathbb{Z}_2$ monopole suppression term,
we observed, for small volumes ($V=4^4$) and on top of the 
strong first order branch of the bulk phase transition  
($\lambda\le 0.7$), tunneling between 
different distinct twist sectors \cite{Barresi:2001dt,Barresi:2002un} 
as well as a dynamical relation between the adjoint Polyakov 
loop and the twist observable, whose non-trivial value signals
the presence of the state $L_A^-=-\frac{1}{3}$ also in this case.
For increasing lattice volume on top of the bulk phase transition 
at low  $\lambda$-values the tunneling between the twist sectors 
becomes more and more suppressed.
Therefore, in agreement with ref. \cite{deForcrand:2002vs},
for small values of the chemical potential the phase transition
seems really first order as we have already argued earlier by monitoring 
other observables such as the average plaquette or the $\mathbb{Z}_2$ 
monopole density. 

\begin{figure}[htb]
\begin{center}
\vspace*{1.0cm}
\includegraphics[angle=0,width=0.48\textwidth]{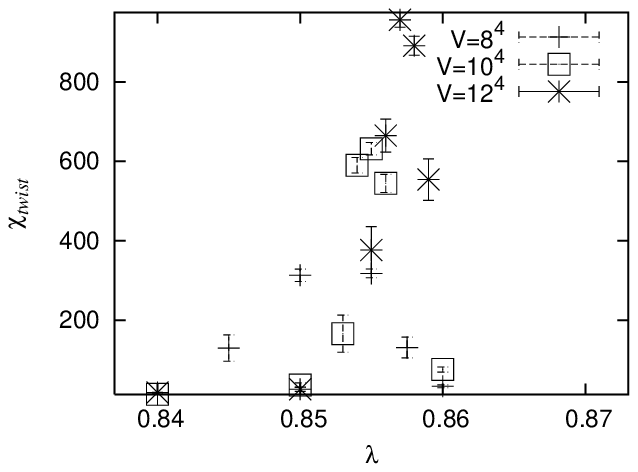}
\includegraphics[angle=0,width=0.48\textwidth]{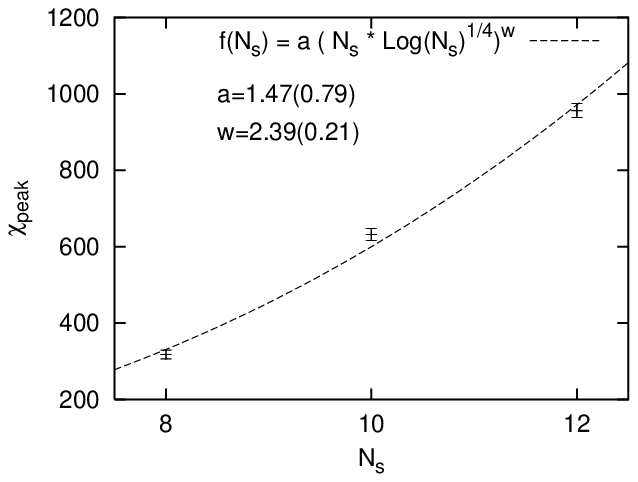}
\end{center}
\caption{L.h.s.: Susceptibility $\chi_{\mathrm{twist}}$ as a function
of $\lambda$ for volumes $8^4, 10^4$ and $12^4$ for $\beta_A=0.65$. R.h.s.:  
Finite-size scaling of the peak values of $\chi_{\mathrm{twist}}$ as
a function of the linear lattice size $N_s$. The dashed line shows the
finite-size scaling function fitted to the data.}
\label{sula085}
\end{figure}

However, as soon as the chemical potential is increased
the situation changes. The observables are no more discontinuous
thus suggesting a very weak first order or second order phase transition.
Indeed, for $\lambda > 0.75$, on top of the transition,
tunneling between the twist sectors is still evident but no suppression is 
seen for increasing volume as Fig. \ref{newtwist} shows
for the lattice size $12^4$. The twists oscillate between -1 and +1 
but the adjoint Polyakov loop fluctuates close to zero and not between +1 and
-1/3, in contrast to what was found in \cite{deForcrand:2002vs}
for $\lambda_V=0$ and in our case at small $\lambda$.
The behaviour of the twist variables reminds that of the fundamental 
Polyakov loop in the $SU(2)$ theory in the fundamental representation,
or of the magnetization in an Ising-like system.
This justifies the definition of an order parameter $<\tilde{z}>$ such as
\begin{eqnarray}
<\tilde{z}> \equiv <\frac{1}{3}(|z_{xt}|+|z_{yt}|+|z_{zt}|)> \, .
\end{eqnarray}
Fig. \ref{sula085} shows the behaviour of the susceptibility of this
new observable $\tilde{z}$
\begin{eqnarray}
\chi_{\mathrm{twist}}=N_S^3
          \cdot(\langle \tilde{z}^2\rangle-\langle\tilde{z}\rangle^2)
\end{eqnarray}
for three different volumes $8^4$ to $12^4$.
\noindent
A finite-size scaling test for the maximum values of the susceptibility 
with the fit-ansatz \cite{Bittner:2002pk}
\begin{eqnarray}
f(N_s) = a (N_s \cdot \log(N_s)^{1/4})^{w}
\end{eqnarray}
shows that the peaks increase with the lattice size with the exponent
$w \simeq 2.4 \pm .2$ which is close to the value of the 4D Ising universality 
class being consistent with the theoretical predictions for the pure
$\mathbb{Z}_2$ monopole system at $\beta_A=0$.
More statistics and larger volumes are needed to confirm this result.
Moreover, the transition does not show any scaling with $N_\tau$, thus
behaving still consistently as a $T=0$ bulk transition.

For larger chemical potential ($\lambda\ge 1.0$)
and positive $\beta_A$, i.e. away from the bulk transition 
(in phase $\mathrm{II}$), tunneling is no more observed and the twist 
sectors become rigid. Keeping the local update algorithm and turning
on the $\mathbb{Z}_2$ monopole potential with 
$\lambda \geq 1$ we can thus
study the theory within the trivial twist sector, i.e. for the case
of a zero number (modulo $2$) of (extended) vortices.

\section{Indication for a phase transition at $T\ne 0$}
\label{sec4}
At finite temperature along a compactified
dimension the periodic b.c. 
on the gauge fields allow for gauge transformations that 
can be classified through 
$\pi_1(SO(3))\sim\mathbb{Z}_2$. In the fundamental representation 
this would indeed translate into a gauge transformation which differs
by a non-trivial element of the center at the boundaries. 
Such transformations can induce a 
selection rule within the Hilbert space. The vacuum
state, and the higher excited states, can exist in
a superposition of the two ($d+1$ for a gauge theory discretized on a
$d$ dimensional torus) different topological states. The creation of a 
vortex in the vacuum,
as measured by the 't Hooft loop, simply means taking the expectation
value between two vacuum states belonging to different topological sectors.
A symmetry breaking argument can thus be applied to the vacuum
state in the Hilbert space, although center symmetry is always unbroken.
The observable $\tilde{z}$ is behaving such that it averages to zero if, at 
finite temperature, the transitions occur among states which exist in a 
superposition of the different topologically distinct eigenstates, while 
it takes a non-zero value otherwise.

We will identify the trivial topological sector, in which we shall 
be particularly interested, with $z_{xt}=z_{yt}=z_{zt}=1$. 
This corresponds to v.e.v. between the (unique even on a torus) 
topologically trivial state, i.e. no creation/annihilation of 
extended vortices (modulo 2).

\begin{figure}[htb]
\begin{center}
\includegraphics[angle=0,width=0.45\textwidth]{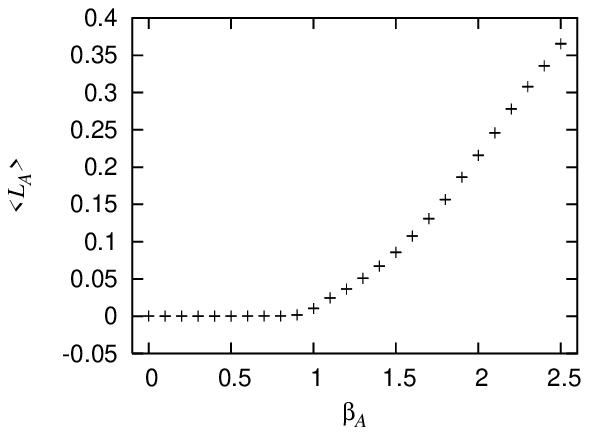}
\includegraphics[angle=0,width=0.45\textwidth]{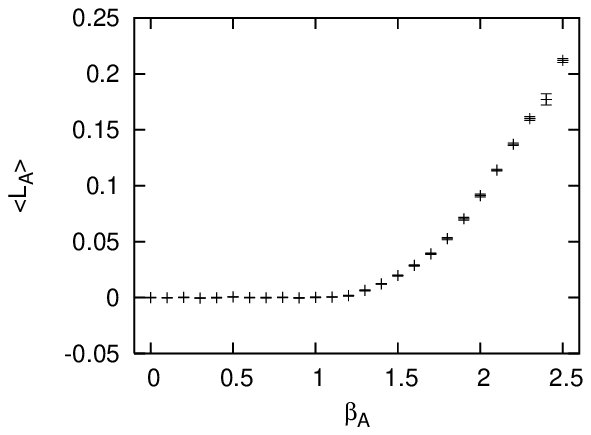}
\end{center}
\caption{Adjoint Polyakov loop $\langle L_A \rangle$ as a function of 
$\beta_A$ in the trivial twist sector for lattice sizes 
$4 \times 16^3$ (l.h.s.) and $6 \times 16^3$ (r.h.s.), both for 
$\lambda=1.0$.}
\label{adjpol}
\end{figure}

\begin{figure}[htb]
\begin{center}
\vspace*{1cm}
\subfigure[$\beta_A=0.9$]{
\includegraphics[angle=0,width=0.23\textwidth]{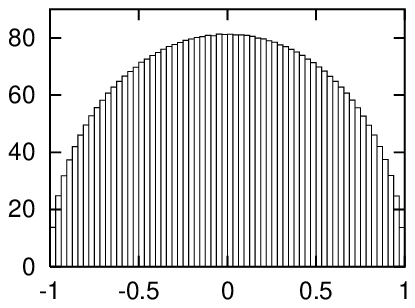}}
\subfigure[$\beta_A=1.1$]{
\includegraphics[angle=0,width=0.23\textwidth]{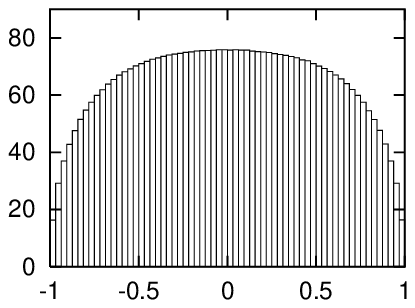}}
\subfigure[$\beta_A=1.2$]{
\includegraphics[angle=0,width=0.23\textwidth]{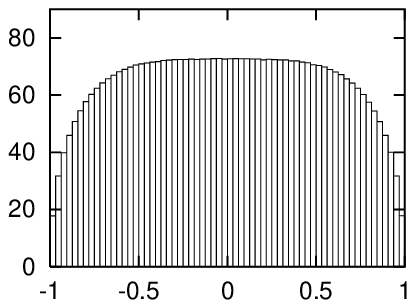}}
\subfigure[$\beta_A=1.4$]{
\includegraphics[angle=0,width=0.23\textwidth]{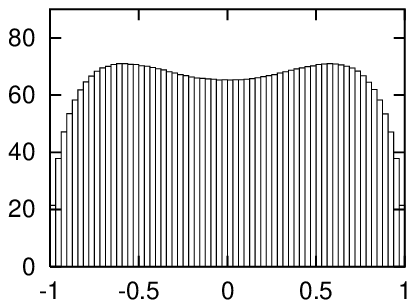}}
\end{center}
\caption{Spatial distribution of the fundamental Polyakov loop 
$\langle L_F \rangle$ in the trivial twist sector at 
$\lambda=1.0$, $V=4\times 16^3$ and at various $\beta_A$-values.}
\label{distr1}
\end{figure}

\begin{figure}[htb]
\begin{center}
\subfigure[$\beta_A=1.4$]{
\includegraphics[angle=0,width=0.23\textwidth]{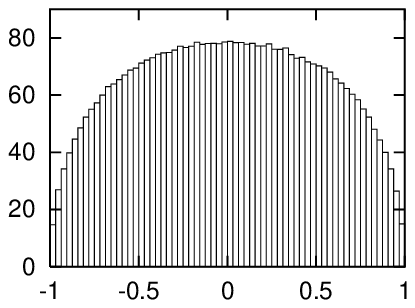}}
\subfigure[$\beta_A=1.5$]{
\includegraphics[angle=0,width=0.23\textwidth]{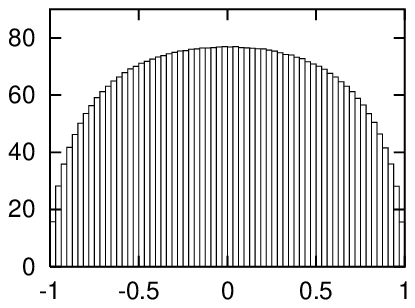}}
\subfigure[$\beta_A=1.6$]{
\includegraphics[angle=0,width=0.23\textwidth]{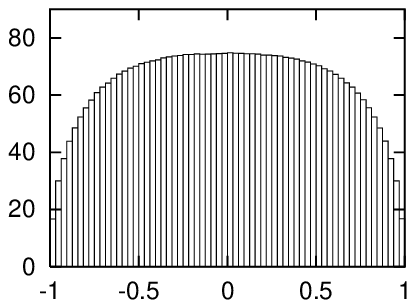}}
\subfigure[$\beta_A=1.8$]{
\includegraphics[angle=0,width=0.23\textwidth]{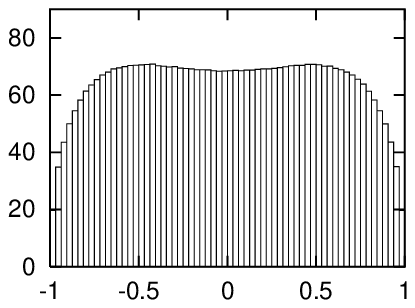}}
\end{center}
\caption{Same as in Fig.~\ref{distr1} for $V=6\times 16^3$.}
\label{distr2} 
\end{figure}

\begin{figure}[h]
\begin{center}
\includegraphics[angle=0,width=0.48\textwidth]{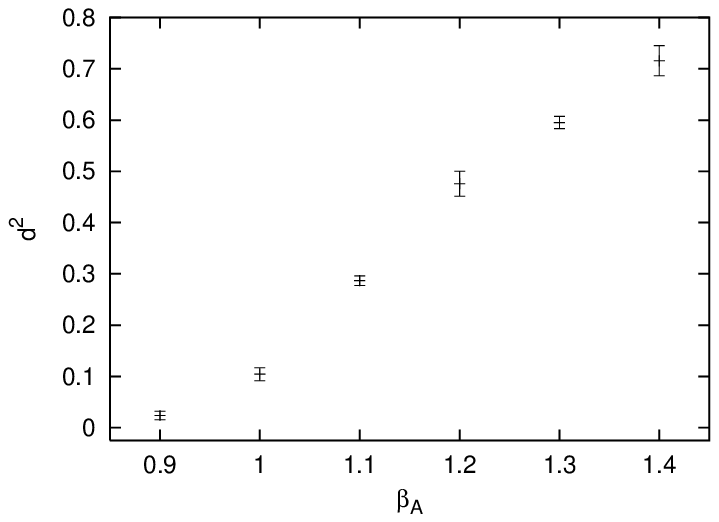}
\includegraphics[angle=0,width=0.48\textwidth]{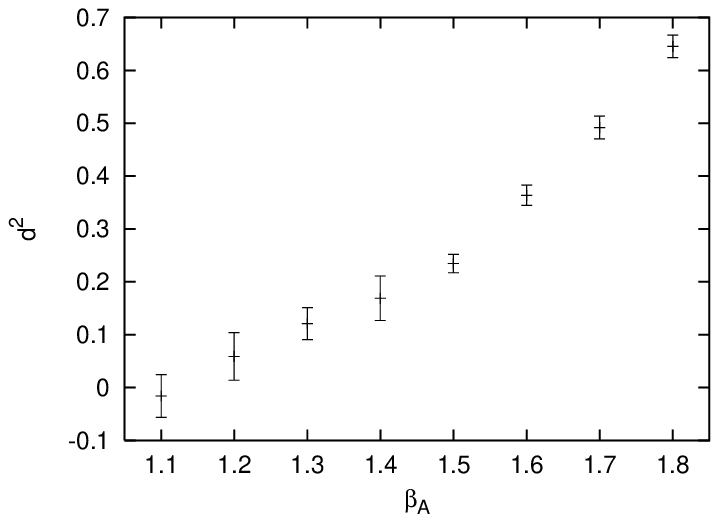}
\end{center}
\caption{Second derivative of the fitting polynomial for the Haar measure
rescaled fundamental Polyakov loop distributions at various $\beta_A$ in the
trivial twist sector ($V=16^3,\lambda=1.0$); the left hand figure for $N_t=4$, 
the right one for $N_t=6$.}
\label{deriv}
\end{figure}

For what we have shown in the previous chapter, we have a local 
way to implement this constraint in the action. 
Indeed, since ergodicity in the $\beta_A-\lambda$ plane is recovered
along the bulk phase transition when it weakens, 
one simply needs to study the system in the broken
phase fixing the trivial twist sector. More than a constraint it actually
simply seems a dynamical feature of the 3+1 dimensional adjoint theory.
Setting $\lambda \ge 1$  and moving parallely  to the $\beta_A$ axis, 
we have studied the phase structure for zero ($N_{\tau} = N_{s}$) 
and non-zero temperature ($N_{\tau} \ll N_{s}$).
The linear spatial lattice size was taken up to $N_{s} \le 16$. We
have determined the distribution of the fundamental Polyakov loop, governing
the expectation value of the adjoint Polyakov loop, as an indicator for 
a physical temperature effect. Appropriate initial conditions were used 
to specify the trivial twist sector monitored throughout the simulations. 
Far enough from  the bulk transition the local updates are keeping 
it fixed. For sake of completeness we have studied the other twist 
sectors as well given the negative value of the Polyakov loop 
associated with these sectors.
For $\lambda \ge 1$ and on the asymmetric lattice ($N_{\tau}=4$) 
a clear finite-$T$ phenomenon appears.  The adjoint Polyakov loop 
still averages to zero for low enough $\beta_A$, while, at higher values
it eventually starts growing, remaining always positive (see Fig. \ref{adjpol}),
consistently with the fixed trivial twist sector and with its interpretation
as the exponential of the free energy of an (infinitely massive) adjoint quark.
For increasing extent of the lattice in the imaginary time direction
the onset of the growth is clearly seen to be shifted to larger $\beta_A$. 
Alternatively one can also consider the spatial distribution of the 
fundamental Polyakov loop. For low $\beta_A$ it shows a broad peak around 
$L_F(\vec{x})=0$ consistent with the Haar measure distribution
$\sim \sqrt{1-L_F^2}$, but with increasing $\beta_A$ 
two symmetric peaks show up, consistent with  $\langle L_A \rangle > 0$
(cf. Fig. \ref{distr1}). The same picture occurs for 
$N_t=6$ but with a shifted coupling as one can see from
Fig. \ref{distr2}.  
In order to highlight this shift somewhat more quantitatively we
fitted these distributions, produced with a statistics of $O(10^4\, \rm {to} 
\, 10^5)$ configurations and rescaled by the Haar measure, 
with some high order polynomials (up to 12th order). 
We verified that the odd derivatives are all zero within the errors 
and that the second derivative departs from zero to positive values 
(see Fig. \ref{deriv}). The corresponding departure points $\beta_A^{0}$
should be taken as lower bounds for the critical values $\beta_A^{crit}$.
We show them in Table I. 
\begin{table}
\begin{center}
\begin{tabular}{|c|c|c|}
\hline
$N_\tau$ & $\beta_A^{0}$ & statistics \\
\hline \hline
4        &  0.92$\pm$ 0.08 & $O(10^5)$   \\
6        &  1.25$\pm$ 0.15 & $O(10^4)$   \\
\hline
\end{tabular}
\end{center}
\caption{The $\beta_A^{0}$-values as explained in the text 
estimated from distributions of the fundamental 
Polyakov loop (see Fig. \ref{deriv}).}
\end{table}
Although we did not determine the critical $\beta_A$-values we see that
the departure points already indicate a reasonable scaling behaviour in $N_t$.
Anyway larger volumes, larger $N_\tau$'s and also other observables will be 
needed to confirm the universality with the fundamental case.
Moreover, although a throughout numerical study is still
required, our data indicate that the point at which the finite 
temperature lines meet the bulk transition line coincides with the 
point where the bulk transition weakens or changes its order 
from $1^{\rm st}$ to $2^{\rm nd}$.
This is consistent with the hypothesis that for large enough 
volume the bulk phase transition would decouple from the 
deconfinement effect and be always weak 1$^{\rm st}$ 
or 2$^{\rm nd}$ order even at $\lambda=0$. 
Unfortunately, a direct numerical verification seems unviable
since from the estimates given in \cite{deForcrand:2002vs} it follows
that it should occur at very large volumes ($V \simeq O(700^4)$ for
the Villain case).

\section{Conclusions}
\label{sec5}
We have carried out a thorough study of pure $SU(2)$
lattice gauge theory in the Wilson adjoint representation, decoup\-ling the strong 
coupling bulk effects from the continuum limit by introducing a chemical 
potential term suppressing $\mathbb{Z}_2$ monopoles. As stressed in
Section \ref{sec2}, our formulation is not - at least not trivially - 
equivalent to the Villain one used in 
\cite{Halliday:1981tm,Datta:1999np,deForcrand:2002vs} 
in the whole parameter range we have explored.
We have moreover included the study of the twist sectors in our analysis,
which had not yet appeared in the literature when the previous studies 
in \cite{Datta:1999np} were carried out. 
In this respect, we have extended the work of 
Ref. \cite{deForcrand:2002vs} to the finite temperature case.
At this point it would be interesting to study also 
the behaviour of the Villain discretization with the $\mathbb{Z}_2$ 
monopole suppression term used in Eq. (\ref{ouraction}).  
By monitoring mainly the twist variables and the adjoint Polyakov loop 
we studied first the character of the bulk transition. The latter 
turned out to become weaker with increasing chemical potential 
$~\lambda~$ turning possibly into a second order transition. 
At strong monopole suppression $~\lambda > 1~$ no bulk transition was found 
anymore. It is this area in the phase diagram where we started to search
for a finite temperature transition. We have found convincing indications
for such a transition at two values of $N_\tau$ from the adjoint Polyakov 
loop and from a typical change
of the distributions of the fundamental Polyakov loop variable. 

\begin{figure}[h]
\begin{center}
\vspace{1.0cm}
\includegraphics[angle=0,width=8.8cm]{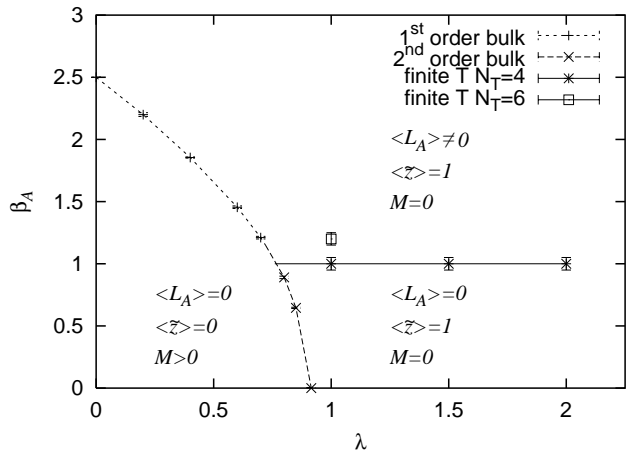}
\end{center}
\caption{Phase diagram in the $\beta_A-\lambda$ plane for $N_\tau=4$ 
(continuous line) and $N_\tau=6$ (single point).}
\label{phdia}
\end{figure}

Due to the strong suppression of tunneling between different twist sectors the 
investigations were carried out with a local update algorithm within the fixed 
trivial twist sector. Of course, a final answer should take into account
also tunnelings between the sectors to be achieved e.g. by multicanonical
updates \cite{Berg:1991cf} or even more promising with parallel 
tempering \cite{Ilgenfritz:2000nj}. 
The results of the present investigation
are collected in Fig. \ref{phdia} where we have also included two further
points for the finite temperature transition seen for $N_t=4$ at larger values
$\lambda =1.5, ~2.0$ in an analogous way as demonstrated for $\lambda=1.0$. 

In the meantime we are carrying out an additional investigation with the  
Pisa disorder parameter (see e.g. \cite{DiGiacomo:2001iy}) 
adapted for the $SO(3)$ case, 
which hopefully will enable us to determine the universality class of the 
transition.  Preliminary data at least support the 
existence of a finite temperature transition decoupled 
from the bulk transition line and at $\beta_A$-values compatible with our
results quoted here. We shall report on this in a forthcoming letter.    

A quantitative study of the observed finite temperature physical transition
could also be viable relying on pure thermodynamic quantities, as already done 
in  \cite{Datta:1999np}, through the Lee-Yang zeros technique or by studying
the free energy of a $\mathbb{Z}_2$ vortex 
\cite{Kovacs:2000sy,deForcrand:2001nd}.
More careful investigations of this kind will hopefully tell us something
about the real nature of the finite temperature transition 
we have reported here. But the mere occurence of such a transition is a 
prerequisite to a still missing proof of universality between 
$SU(2)$ and $SO(3)$ lattice gauge theories. 

\section*{Acknowledgments}
We would like to thank V. Bornyakov, M. Creutz, A. Di Giacomo, 
H. Ichie, E.-M. Ilgenfritz , R. Kenna, T. Kovacs, M. Peardon, 
and M. Pepe for helpful comments and discussions. 
Special thanks go to  Oliver Jahn, Philippe de Forcrand and Terry Tomboulis 
for detailed critical remarks. 
This work was financially supported by a EU-TMR network under the contract
FMRX-CT97-0122 and by the DFG-funded graduate school GK 271.

\bibliography{bib.bib}

\begin{thebibliography}{36}
\expandafter\ifx\csname natexlab\endcsname\relax\def\natexlab#1{#1}\fi
\expandafter\ifx\csname bibnamefont\endcsname\relax
  \def\bibnamefont#1{#1}\fi
\expandafter\ifx\csname bibfnamefont\endcsname\relax
  \def\bibfnamefont#1{#1}\fi
\expandafter\ifx\csname citenamefont\endcsname\relax
  \def\citenamefont#1{#1}\fi
\expandafter\ifx\csname url\endcsname\relax
  \def\url#1{\texttt{#1}}\fi
\expandafter\ifx\csname urlprefix\endcsname\relax\def\urlprefix{URL }\fi
\providecommand{\bibinfo}[2]{#2}
\providecommand{\eprint}[2][]{\url{#2}}

\bibitem[{\citenamefont{McLerran and Svetitsky}(1981)}]{McLerran:1981pk}
\bibinfo{author}{\bibfnamefont{L.~D.} \bibnamefont{McLerran}} \bibnamefont{and}
  \bibinfo{author}{\bibfnamefont{B.}~\bibnamefont{Svetitsky}},
  \bibinfo{journal}{Phys. Lett.} \textbf{\bibinfo{volume}{B98}},
  \bibinfo{pages}{195} (\bibinfo{year}{1981}).

\bibitem[{\citenamefont{Kuti et~al.}(1981)\citenamefont{Kuti, Polonyi, and
  Szlachanyi}}]{Kuti:1981gh}
\bibinfo{author}{\bibfnamefont{J.}~\bibnamefont{Kuti}},
  \bibinfo{author}{\bibfnamefont{J.}~\bibnamefont{Polonyi}}, \bibnamefont{and}
  \bibinfo{author}{\bibfnamefont{K.}~\bibnamefont{Szlachanyi}},
  \bibinfo{journal}{Phys. Lett.} \textbf{\bibinfo{volume}{B98}},
  \bibinfo{pages}{199} (\bibinfo{year}{1981}).

\bibitem[{\citenamefont{Polyakov}(1978)}]{Polyakov:1978vu}
\bibinfo{author}{\bibfnamefont{A.~M.} \bibnamefont{Polyakov}},
  \bibinfo{journal}{Phys. Lett.} \textbf{\bibinfo{volume}{B72}},
  \bibinfo{pages}{477} (\bibinfo{year}{1978}).

\bibitem[{\citenamefont{Susskind}(1979)}]{Susskind:1979up}
\bibinfo{author}{\bibfnamefont{L.}~\bibnamefont{Susskind}},
  \bibinfo{journal}{Phys. Rev.} \textbf{\bibinfo{volume}{D20}},
  \bibinfo{pages}{2610} (\bibinfo{year}{1979}).

\bibitem[{\citenamefont{Svetitsky and Yaffe}(1982)}]{Svetitsky:1982gs}
\bibinfo{author}{\bibfnamefont{B.}~\bibnamefont{Svetitsky}} \bibnamefont{and}
  \bibinfo{author}{\bibfnamefont{L.~G.} \bibnamefont{Yaffe}},
  \bibinfo{journal}{Nucl. Phys.} \textbf{\bibinfo{volume}{B210}},
  \bibinfo{pages}{423} (\bibinfo{year}{1982}).

\bibitem[{\citenamefont{Fortunato et~al.}(2001)\citenamefont{Fortunato, Karsch,
  Petreczky, and Satz}}]{Fortunato:2000vf}
\bibinfo{author}{\bibfnamefont{S.}~\bibnamefont{Fortunato}},
  \bibinfo{author}{\bibfnamefont{F.}~\bibnamefont{Karsch}},
  \bibinfo{author}{\bibfnamefont{P.}~\bibnamefont{Petreczky}},
  \bibnamefont{and} \bibinfo{author}{\bibfnamefont{H.}~\bibnamefont{Satz}},
  \bibinfo{journal}{Phys. Lett.} \textbf{\bibinfo{volume}{B502}},
  \bibinfo{pages}{321} (\bibinfo{year}{2001}), \eprint{hep-lat/0011084}.

\bibitem[{\citenamefont{'t~Hooft}(1979)}]{'tHooft:1979uj}
\bibinfo{author}{\bibfnamefont{G.}~\bibnamefont{'t~Hooft}},
  \bibinfo{journal}{Nucl. Phys.} \textbf{\bibinfo{volume}{B153}},
  \bibinfo{pages}{141} (\bibinfo{year}{1979}).

\bibitem[{\citenamefont{Seiberg and
  Witten}(1994{\natexlab{a}})}]{Seiberg:1994rs}
\bibinfo{author}{\bibfnamefont{N.}~\bibnamefont{Seiberg}} \bibnamefont{and}
  \bibinfo{author}{\bibfnamefont{E.}~\bibnamefont{Witten}},
  \bibinfo{journal}{Nucl. Phys.} \textbf{\bibinfo{volume}{B426}},
  \bibinfo{pages}{19} (\bibinfo{year}{1994}{\natexlab{a}}),
  \eprint{hep-th/9407087}.

\bibitem[{\citenamefont{Seiberg and
  Witten}(1994{\natexlab{b}})}]{Seiberg:1994aj}
\bibinfo{author}{\bibfnamefont{N.}~\bibnamefont{Seiberg}} \bibnamefont{and}
  \bibinfo{author}{\bibfnamefont{E.}~\bibnamefont{Witten}},
  \bibinfo{journal}{Nucl. Phys.} \textbf{\bibinfo{volume}{B431}},
  \bibinfo{pages}{484} (\bibinfo{year}{1994}{\natexlab{b}}),
  \eprint{hep-th/9408099}.

\bibitem[{\citenamefont{Polyakov}(1977)}]{Polyakov:1977fu}
\bibinfo{author}{\bibfnamefont{A.~M.} \bibnamefont{Polyakov}},
  \bibinfo{journal}{Nucl. Phys.} \textbf{\bibinfo{volume}{B120}},
  \bibinfo{pages}{429} (\bibinfo{year}{1977}).

\bibitem[{\citenamefont{Holland et~al.}(2003)\citenamefont{Holland, Minkowski,
  Pepe, and Wiese}}]{Holland:2003jy}
\bibinfo{author}{\bibfnamefont{K.}~\bibnamefont{Holland}},
  \bibinfo{author}{\bibfnamefont{P.}~\bibnamefont{Minkowski}},
  \bibinfo{author}{\bibfnamefont{M.}~\bibnamefont{Pepe}}, \bibnamefont{and}
  \bibinfo{author}{\bibfnamefont{U.~J.} \bibnamefont{Wiese}}
  (\bibinfo{year}{2003}), \eprint{hep-lat/0302023}.

\bibitem[{\citenamefont{Michael}(1985)}]{Michael:1985ne}
\bibinfo{author}{\bibfnamefont{C.}~\bibnamefont{Michael}},
  \bibinfo{journal}{Nucl. Phys.} \textbf{\bibinfo{volume}{B259}},
  \bibinfo{pages}{58} (\bibinfo{year}{1985}).

\bibitem[{\citenamefont{de~Forcrand and Philipsen}(2000)}]{deForcrand:1999kr}
\bibinfo{author}{\bibfnamefont{P.}~\bibnamefont{de~Forcrand}} \bibnamefont{and}
  \bibinfo{author}{\bibfnamefont{O.}~\bibnamefont{Philipsen}},
  \bibinfo{journal}{Phys. Lett.} \textbf{\bibinfo{volume}{B475}},
  \bibinfo{pages}{280} (\bibinfo{year}{2000}), \eprint{hep-lat/9912050}.

\bibitem[{\citenamefont{Bhanot and Creutz}(1981)}]{Bhanot:1981eb}
\bibinfo{author}{\bibfnamefont{G.}~\bibnamefont{Bhanot}} \bibnamefont{and}
  \bibinfo{author}{\bibfnamefont{M.}~\bibnamefont{Creutz}},
  \bibinfo{journal}{Phys. Rev.} \textbf{\bibinfo{volume}{D24}},
  \bibinfo{pages}{3212} (\bibinfo{year}{1981}).

\bibitem[{\citenamefont{Greensite and Lautrup}(1981)}]{Greensite:1981hw}
\bibinfo{author}{\bibfnamefont{J.}~\bibnamefont{Greensite}} \bibnamefont{and}
  \bibinfo{author}{\bibfnamefont{B.}~\bibnamefont{Lautrup}},
  \bibinfo{journal}{Phys. Rev. Lett.} \textbf{\bibinfo{volume}{47}},
  \bibinfo{pages}{9} (\bibinfo{year}{1981}).

\bibitem[{\citenamefont{Creutz and Moriarty}(1982)}]{Creutz:1982ga}
\bibinfo{author}{\bibfnamefont{M.}~\bibnamefont{Creutz}} \bibnamefont{and}
  \bibinfo{author}{\bibfnamefont{K.~J.~M.} \bibnamefont{Moriarty}},
  \bibinfo{journal}{Nucl. Phys.} \textbf{\bibinfo{volume}{B210}},
  \bibinfo{pages}{50} (\bibinfo{year}{1982}).

\bibitem[{\citenamefont{Halliday and
  Schwimmer}(1981{\natexlab{a}})}]{Halliday:1981te}
\bibinfo{author}{\bibfnamefont{I.~G.} \bibnamefont{Halliday}} \bibnamefont{and}
  \bibinfo{author}{\bibfnamefont{A.}~\bibnamefont{Schwimmer}},
  \bibinfo{journal}{Phys. Lett.} \textbf{\bibinfo{volume}{B101}},
  \bibinfo{pages}{327} (\bibinfo{year}{1981}{\natexlab{a}}).

\bibitem[{\citenamefont{Caneschi et~al.}(1982)\citenamefont{Caneschi, Halliday,
  and Schwimmer}}]{Caneschi:1982ik}
\bibinfo{author}{\bibfnamefont{L.}~\bibnamefont{Caneschi}},
  \bibinfo{author}{\bibfnamefont{I.~G.} \bibnamefont{Halliday}},
  \bibnamefont{and}
  \bibinfo{author}{\bibfnamefont{A.}~\bibnamefont{Schwimmer}},
  \bibinfo{journal}{Nucl. Phys.} \textbf{\bibinfo{volume}{B200}},
  \bibinfo{pages}{409} (\bibinfo{year}{1982}).

\bibitem[{\citenamefont{de~Forcrand and von Smekal}(2002)}]{deForcrand:2001nd}
\bibinfo{author}{\bibfnamefont{P.}~\bibnamefont{de~Forcrand}} \bibnamefont{and}
  \bibinfo{author}{\bibfnamefont{L.}~\bibnamefont{von Smekal}},
  \bibinfo{journal}{Phys. Rev.} \textbf{\bibinfo{volume}{D66}},
  \bibinfo{pages}{011504} (\bibinfo{year}{2002}), \eprint{hep-lat/0107018}.

\bibitem[{\citenamefont{de~Forcrand and Jahn}(2003)}]{deForcrand:2002vs}
\bibinfo{author}{\bibfnamefont{P.}~\bibnamefont{de~Forcrand}} \bibnamefont{and}
  \bibinfo{author}{\bibfnamefont{O.}~\bibnamefont{Jahn}},
  \bibinfo{journal}{Nucl. Phys.} \textbf{\bibinfo{volume}{B651}},
  \bibinfo{pages}{125} (\bibinfo{year}{2003}), \eprint{hep-lat/0211004}.

\bibitem[{\citenamefont{Datta and Gavai}(1997)}]{Datta:1998nv}
\bibinfo{author}{\bibfnamefont{S.}~\bibnamefont{Datta}} \bibnamefont{and}
  \bibinfo{author}{\bibfnamefont{R.~V.} \bibnamefont{Gavai}},
  \bibinfo{journal}{Phys. Rev.} \textbf{\bibinfo{volume}{D57}},
  \bibinfo{pages}{6618} (\bibinfo{year}{1998}), \eprint{hep-lat/9708026}.

\bibitem[{\citenamefont{Datta and Gavai}(1999)}]{Datta:1999np}
\bibinfo{author}{\bibfnamefont{S.}~\bibnamefont{Datta}} \bibnamefont{and}
  \bibinfo{author}{\bibfnamefont{R.~V.} \bibnamefont{Gavai}},
  \bibinfo{journal}{Phys. Rev.} \textbf{\bibinfo{volume}{D60}},
  \bibinfo{pages}{034505} (\bibinfo{year}{1999}),
  \eprint[http://arXiv.org/abs]{hep-lat/9901006}.

\bibitem[{\citenamefont{Datta and Gavai}(2000)}]{Datta:1999ep}
\bibinfo{author}{\bibfnamefont{S.}~\bibnamefont{Datta}} \bibnamefont{and}
  \bibinfo{author}{\bibfnamefont{R.~V.} \bibnamefont{Gavai}},
  \bibinfo{journal}{Phys. Rev.} \textbf{\bibinfo{volume}{D62}},
  \bibinfo{pages}{054512} (\bibinfo{year}{2000}), \eprint{hep-lat/9909139}.

\bibitem[{\citenamefont{Barresi
  et~al.}(2002{\natexlab{a}})\citenamefont{Barresi, Burgio, and
  Muller-Preussker}}]{Barresi:2001dt}
\bibinfo{author}{\bibfnamefont{A.}~\bibnamefont{Barresi}},
  \bibinfo{author}{\bibfnamefont{G.}~\bibnamefont{Burgio}}, \bibnamefont{and}
  \bibinfo{author}{\bibfnamefont{M.}~\bibnamefont{M\"uller-Preussker}},
  \bibinfo{journal}{Nucl. Phys. Proc. Suppl.} \textbf{\bibinfo{volume}{106}},
  \bibinfo{pages}{495} (\bibinfo{year}{2002}{\natexlab{a}}),
  \eprint{hep-lat/0110139}.

\bibitem[{\citenamefont{Barresi
  et~al.}(2002{\natexlab{b}})\citenamefont{Barresi, Burgio, and
  Muller-Preussker}}]{Barresi:2002un}
\bibinfo{author}{\bibfnamefont{A.}~\bibnamefont{Barresi}},
  \bibinfo{author}{\bibfnamefont{G.}~\bibnamefont{Burgio}}, \bibnamefont{and}
  \bibinfo{author}{\bibfnamefont{M.}~\bibnamefont{M\"uller-Preussker}},
  \bibinfo{journal}{Nucl. Phys. Proc. Suppl.} \textbf{\bibinfo{volume}{119}},
  \bibinfo{pages}{571} (\bibinfo{year}{2003}{\natexlab{b}}),
  \eprint{hep-lat/0209011}.

\bibitem[{\citenamefont{Halliday and
  Schwimmer}(1981{\natexlab{b}})}]{Halliday:1981tm}
\bibinfo{author}{\bibfnamefont{I.~G.} \bibnamefont{Halliday}} \bibnamefont{and}
  \bibinfo{author}{\bibfnamefont{A.}~\bibnamefont{Schwimmer}},
  \bibinfo{journal}{Phys. Lett.} \textbf{\bibinfo{volume}{B102}},
  \bibinfo{pages}{337} (\bibinfo{year}{1981}{\natexlab{b}}).

\bibitem[{\citenamefont{Cheluvaraja and Sharathchandra}(1996)}]{Cheluvaraja:1996zn}
\bibinfo{author}{\bibfnamefont{S.}~\bibnamefont{Cheluvaraja}} \bibnamefont{and}
  \bibinfo{author}{\bibfnamefont{H. S.}~\bibnamefont{Sharathchandra}}
  (\bibinfo{year}{1996}), \eprint{hep-lat/9611001}.

\bibitem[{\citenamefont{Bittner et~al.}(2002)}]{Bittner:2002pk}
\bibinfo{author}{\bibfnamefont{E.}~\bibnamefont{Bittner}}, 
  \bibinfo{author}{\bibfnamefont{W.}~\bibnamefont{Janke}} \bibnamefont{and}
  \bibinfo{author}{\bibfnamefont{H.} \bibnamefont{Markum}},
  \bibinfo{journal}{Phys. Rev.} \textbf{\bibinfo{volume}{D66}}, 
  \bibinfo{pages}{024008} (\bibinfo{year}{2002}),
  \eprint{hep-lat/0205023}.

\bibitem[{\citenamefont{Berg and Neuhaus}(1991)}]{Berg:1991cf}
\bibinfo{author}{\bibfnamefont{B.~A.} \bibnamefont{Berg}} \bibnamefont{and}
  \bibinfo{author}{\bibfnamefont{T.}~\bibnamefont{Neuhaus}},
  \bibinfo{journal}{Phys. Lett.} \textbf{\bibinfo{volume}{B267}},
  \bibinfo{pages}{249} (\bibinfo{year}{1991}).

\bibitem[{\citenamefont{Ilgenfritz et~al.}(2000)\citenamefont{Ilgenfritz,
  Kerler, Muller-Preussker, and St\"uben}}]{Ilgenfritz:2000nj}
  \bibinfo{author}{\bibfnamefont{E.~M.} \bibnamefont{Ilgenfritz}},
  \bibinfo{author}{\bibfnamefont{W.}~\bibnamefont{Kerler}},
  \bibinfo{author}{\bibfnamefont{M.}~\bibnamefont{M\"uller-Preussker}},
  \bibnamefont{and} \bibinfo{author}{\bibfnamefont{H.}~\bibnamefont{St\"uben}},
  \bibinfo{journal}{Phys. Rev.} \textbf{\bibinfo{volume}{D65}}, 
  \bibinfo{pages}{094506} (\bibinfo{year}{2002}),
  \eprint{hep-lat/0007039}.

\bibitem[{\citenamefont{Di Giacomo}(2000)}]{DiGiacomo:2001iy}
\bibinfo{author}{\bibfnamefont{A.} \bibnamefont{Di Giacomo}},
  \bibinfo{journal}{Nucl. Phys.} \textbf{\bibinfo{volume}{A702}},
  \bibinfo{pages}{73} (\bibinfo{year}{2002}), \eprint{hep-lat/0112002}. 

\bibitem[{\citenamefont{Kovacs and Tomboulis}(2000)}]{Kovacs:2000sy}
\bibinfo{author}{\bibfnamefont{T.} \bibnamefont{Kovacs}} \bibnamefont{and}
  \bibinfo{author}{\bibfnamefont{T.}~\bibnamefont{Tomboulis}},
  \bibinfo{journal}{Phys. Rev. Lett.} \textbf{\bibinfo{volume}{85}},
  \bibinfo{pages}{704} (\bibinfo{year}{2000}), \eprint{hep-lat/0002004}.
  
\end{thebibliography}
\end{document}